\documentclass[aps,prd,preprint,nofootinbib,longbibliography]{revtex4-1}
\usepackage{blindtext}
\usepackage{mathtools}
\usepackage{xcolor}
\usepackage{amssymb,amsmath,bbold}
\usepackage{physics}
\usepackage{dsfont}
\usepackage{colortbl}

\usepackage[utf8]{inputenc}
\usepackage[english]{babel}
\usepackage{hyperref}
\newcommand{\Bra}[2]{\mathopen{_{#2}}\bra{#1}}
\newcommand{\Ket}[2]{\ket{#1}\mathclose{_{#2}}}

\newcommand{\normord}[1]{\mathopen{:\,}#1\mathclose{:}}

\begin{document}
\title{Frequency Spectra Analysis of Space and Time Averaged Quantum Stress Tensor Fluctuations} \author{Peter Wu$^{1\dagger}$} \author{L. H. Ford$^{1\ddagger}$} \author{Enrico D. Schiappacasse$^{2*}$}
\affiliation{$^1$Institute of Cosmology, Department of Physics and Astronomy, Tufts University, Medford, Massachusetts 02155, USA\\
$^2$Department of Physics and Astronomy \& Rice University Academy of Fellows, Rice University, Houston, Texas, 77005, USA}

\begin{abstract}
Observing physical effects of large quantum stress tensor fluctuations requires knowledge of the interactions between the probe and the particles of the underlying quantum fields.  The quantum stress tensor operators must first be averaged in time alone or space and time to confer meaningful results, the details of which may correspond to the physical measurement process.  We build on prior results to characterize the particle frequencies associated with quantum fluctuations of different magnitudes.  For the square of time derivatives of the massless scalar field in a spherical cavity, we find that these frequencies are bounded above in a power law behavior.  Our findings provide a way identify the largest quantum fluctuation that may be probed in experiments relying on frequency-dependent interactions.
\end{abstract}

\maketitle

\section{Introduction}
As the search for a full theory of quantum gravity continues, extensions to the semiclassical theory of gravity remain a promising endeavor for understanding quantum gravitational effects.  In the semiclassical theory of gravity, the stress-energy-momentum tensor is treated as a quantum operator, related to the classical Einstein tensor by taking an expectation value~\cite{Ford:2005qz}.  This approach has seen success in a number of scenarios, ranging from Hawking radiation from black holes~\cite{Hawking} and the resulting gravitational backreaction~\cite{birrell_davies_1982} to quantum particle creation~\cite{Ford_2021}.  Extensions to the semiclassical theory include the addition of higher order derivatives in the metric tensor~\cite{PhysRevD.17.414}, with consequences that include the production of gravitons in the early universe~\cite{Schiappacasse:2016nei,PhysRevD.94.063517}, and the incorporation of fluctuations of the quantum stress tensor around its mean value~\cite{Hu2008}.  The latter approach motivates the study of quantum fluctuations of stress tensor operators, which have been shown to source a number of possible effects, such as geodesic focusing~\cite{Borgman:2003dm} and imprints on power spectra~\cite{Ford:2010wd,PhysRevD.95.063524}.  

In recent years, focus has shifted to large vacuum fluctuations due to hints that they may be observable in experiments.  A generic property of normal ordered operators quadratic in the fields is the divergence of the higher moments, a problem formally addressed by averaging these operators in time alone or space and time.  Physically, we may interpret this averaging as encoding the details of experimental measurements, though the correspondence between the two remains under investigation.  In two-dimensional conformal field theory with a Gaussian temporal sampling function, the probability~$P(x)$ of measuring a fluctuation of magnitude~$x$ is a shifted Gamma distribution~\cite{Fewster:2010mc,PhysRevD.101.025010,Fewster2019} bounded below by the optimal quantum inequality bound~\cite{Fewster:2004nj}.  In four dimensions, the situation is qualitatively similar, and studies have been conducted for time averaging~\cite{Fewster:2012ej,Fewster:2015hga} and spacetime averaging~\cite{PhysRevD.101.025006}.  These results, numerically verified in Refs.~\cite{Schiappacasse:2017oqu,PhysRevD.103.125014}, suggest the probability distribution~$P(x)$ asymptotically falls more slowly with~$x$ than a decaying exponential function.  We will be particularly interested in the conclusions of Refs.~\cite{Fewster:2015hga,PhysRevD.101.025006}, as they assume smooth, compactly supported sampling functions.  These are functions that are strictly zero outside a finite, bounded domain, and may better reflect the reality that physical measurements necessarily take place in a finite spacetime region.

These findings suggest the probabilities of measuring large vacuum fluctuations of stress tensor operators may not be as negligible as one might have expected, spurring work into possible physical effects such as focusing of geodesics induced by spacetime curvature fluctuations~\cite{PhysRevLett.107.021303,PhysRevD.102.126018} and enhanced barrier penetration rates of charged particles due to radiation pressure fluctuations~\cite{Huang:2016kmx}.  Related phenomena in other systems that emerge from similar quadratic operators are also expected to occur: false vacuum decay of a self-interacting scalar field may be dominated by a pathway~\cite{PhysRevD.105.085025} different from the usual instanton approximation~\cite{PhysRevD.15.2929}, Rydberg atoms may exhibit velocity fluctuations in response to a sequence of short laser pulses~\cite{PhysRevA.104.012208}, and low-temperature light scattering experiments may find large variations in the number of scattered photons~\cite{PhysRevResearch.2.032028}.

In this paper, we investigate a different but related question.  Models proposing experimental effects inevitably assume certain interactions between the probe and the particles of the quantum fields, and these interactions are often dependent on the particle frequencies.  The relationship between the magnitudes of quantum fluctuations and the angular frequencies of the constituent particles remains unclear, and better understanding will help produce more accurate models for future experiments.  In Sec.~\ref{sec:background}, we review a numerical approach for diagonalizing bosonic operators that are quadratic in the fields, a category that includes stress tensor operators.  In Sec.~\ref{sec:theory}, we build on previous results to investigate the relationship between the frequencies of the particle contents and the magnitudes of the fluctuations.  In Sec.~\ref{sec:results}, we discuss numerical simulations that confirm the analytical calculations and, in doing so, provide estimates of constants that are not well predicted analytically.  In Sec.~\ref{sec:application}, we briefly describe a physical application as an example showing how these results may be of interest in experimental contexts.  Finally, in Sec.~\ref{sec:conclusion}, we summarize our findings and consider some future investigations.

Units in which the reduced Planck constant~$\hbar$ and the speed of light are set to unity are used throughout this paper.

\section{Numerical approach to spacetime-averaged quadratic operators}
\label{sec:background}
Components of the normal ordered stress-energy-momentum tensor operator, such as those of the massless scalar field, may be generally expanded in terms of creation and annihilation operators as
\begin{equation}
\mathcal{T}(t,\mathbf{r}) = \frac{1}{2}\sum_{\mathbf{k}\mathbf{k}'} \Big[2a^\dagger_{\mathbf{k}}a_{\mathbf{k}'} F_{\mathbf{k}\mathbf{k}'}(t,\mathbf{r}) + a_{\mathbf{k}}a_{\mathbf{k}'} G_{\mathbf{k}\mathbf{k}'} (t,\mathbf{r}) +a^\dagger_{\mathbf{k}}a^\dagger_{\mathbf{k}'}G^*_{\mathbf{k}\mathbf{k}'} (t,\mathbf{r})\Big]\,, 
\label{eq:quadgeneral}
\end{equation}
where
\begin{equation}
[a_{\mathbf{k}},a^\dagger_{\mathbf{k}'}]=\delta_{\mathbf{k}\mathbf{k}'}\mathbb{1}\,\,\,\text{and}\,\,\,[a_{\mathbf{k}},a_{\mathbf{k}'}]=[a^\dagger_{\mathbf{k}},a^\dagger_{\mathbf{k}'}]=0\,.\label{eq:com}
\end{equation}
Here~$F_{\mathbf{k}\mathbf{k}'}(t,\mathbf{r})$ and~$G_{\mathbf{k}\mathbf{k}'}(t,\mathbf{r})$ are matrix elements that depend on the specific choice of operator, and~$\mathbb{1}$ is the identity operator.  Because physical measurements take place in a finite region of spacetime, a more meaningful quantity to consider is the operator~$\mathcal{T}(t,\mathbf{r})$ averaged in space and time with compactly supported sampling functions,~$g(\mathbf{r})$ and~$f(t)$, respectively, giving
\begin{equation}
    \bar{\mathcal{T}}\equiv\int_{-\infty}^\infty dt \,f(t) \int_{\mathcal{V}} d^3r \,g(\mathbf{r})\,\mathcal{T}(t,\mathbf{r})\,.
\end{equation}
We further assume the spatial sampling function~$g(\mathbf{r})$ is non-negative, spherically symmetric, and real, with unit integral over all space.  Likewise, the temporal sampling function~$f(t)$ is non-negative, even, and real, with unit integral over all time.  Although experimental data is lacking, we expect the details of these functions to correlate with factors in the measurement process.  The Fourier transforms of these sampling functions are defined as
\begin{equation}
    \hat{g}(\mathbf{k}) = \int_{\mathcal{V}} d^3r \, g(\mathbf{r}) e^{-i\mathbf{k}\cdot\mathbf{r}}
\end{equation}
and
\begin{equation}
    \hat{f}(\omega)=\int_{-\infty}^\infty  dt \,f(t) e^{-i\omega t}\,.
\end{equation}
We may write the analogous expression to Eq.~(\ref{eq:quadgeneral}) for the spacetime-averaged operator~$\bar{\mathcal{T}}$ as
\begin{equation}
\bar{\mathcal{T}} = \frac{1}{2}\sum_{\mathbf{k}\mathbf{k}'} \Big[2a^\dagger_{\mathbf{k}}a_{\mathbf{k}'} \bar{F}_{\mathbf{k}\mathbf{k}'}+ a_{\mathbf{k}}a_{\mathbf{k}'} \bar{G}_{\mathbf{k}\mathbf{k}'} +a^\dagger_{\mathbf{k}}a^\dagger_{\mathbf{k}'}\bar{G}^*_{\mathbf{k}\mathbf{k}'} \Big]\,.
\label{eq:operator}
\end{equation}
As we are concerned with vacuum fluctuations of these quadratic operators, let us consider the Minkowski vacuum state~$\Ket{\mathbf{0}}{a}$, defined as the state where~$a_{\mathbf{k}}\Ket{\mathbf{0}}{a}=0$ for all~$\mathbf{k}$.  We immediately see that the vacuum state is not an eigenstate of the operator~$\bar{\mathcal{T}}$, so quantum fluctuations will be present in the vacuum state.

To perform numerical simulations of these vacuum fluctuations, we need to characterize the eigenstates of the operator~$\bar{\mathcal{T}}$.  We follow a numerical method developed in Refs.~\cite{Dawson,Schiappacasse:2017oqu}, briefly summarized here.  For the cases we consider, the matrix elements of~$\bar{F}$ and~$\bar{G}$ are real, i.e.,~$\bar{G}_{\mathbf{k}\mathbf{k}'}=\bar{G}_{\mathbf{k}\mathbf{k}'}^*$.  A Bogoliubov transformation~\cite{Bogolyubov:1947zz} relates the original set of creation and annihilation operators~$\{a_{\mathbf{k}}^\dagger,a_{\mathbf{k}}\}$ to a new set~$\{b_{\mathbf{k}}^\dagger,b_{\mathbf{k}}\}$.  The two sets of operators are related by the linear transformation
\begin{equation}
a_{\mathbf{k}} = \sum_{\mathbf{k}'} \left(A_{\mathbf{k}\mathbf{k}'} b_{\mathbf{k}'} +B_{\mathbf{k}\mathbf{k}'} b^\dagger_{\mathbf{k}'}\right)\,.
\label{eq:bogoliubov}
\end{equation}
The~$A$ and~$B$ matrices contain the real Bogoliubov coefficients that transform $\{a_{\mathbf{k}}^\dagger,a_{\mathbf{k}}\}$ to~$\{b_{\mathbf{k}}^\dagger,b_{\mathbf{k}}\}$.  The new set of creation and annihilation operators~$\{b_{\mathbf{k}}^\dagger,b_{\mathbf{k}}\}$ acts on their own respective particle number states
\begin{equation}
\Ket{\mathbf{m}}{b} \equiv \Ket{m_{\mathbf{k}_1},m_{\mathbf{k}_2},m_{\mathbf{k}_3},\dotsm}{b}\,.
\label{eq:bket}
\end{equation}
Note that the physical particles of interest are the~$a$ particles of the Minkowski vacuum state~$\Ket{\mathbf{0}}{a}$, not the~$b$ particles that emerge as a consequence of the Bogoliubov transformation.  That is to say, any observable physics arising from vacuum fluctuations comes from interactions with the real~$a$ particles.  Substituting Eq.~(\ref{eq:bogoliubov}) into Eq.~(\ref{eq:operator}), we may rewrite~$\bar{\mathcal{T}}$ in the diagonal form
\begin{equation}
\bar{\mathcal{T}} = \sum_{\mathbf{k}} \lambda_{\mathbf{k}} b^\dagger_{\mathbf{k}}b_{\mathbf{k}} + C_{\text{shift}} \mathbb{1}\,,
\label{eq:diagonal}
\end{equation}
where~$\mathbb{1}$ is again the identity operator.  Here~$\{\lambda_{\mathbf{k}}\}$ and~$C_{\text{shift}}$ are constants that depend on the matrix elements of $\bar{F}$ and~$\bar{G}$.  These constants may be derived by mandating~$\{b_{\mathbf{k}}^\dagger,b_{\mathbf{k}}\}$ obey the usual commutation relations analogous to those in Eq.~(\ref{eq:com}), in addition to requiring the diagonalization conditions leading to Eq.~(\ref{eq:diagonal}).  Theoretical considerations from quantum inequalities bound the eigenvalues of~$\bar{\mathcal{T}}$ from below, so~$\lambda_{\mathbf{k}}\geq0$ for all~$\mathbf{k}$ and~$C_{\text{shift}}<0$.  We interpret the~$b$ particle number states~$\Ket{\mathbf{m}}{b}$ as the eigenstates of the operator~$\bar{\mathcal{T}}$.  The eigenvalues~$x$ are calculated by acting~$\bar{\mathcal{T}}$ on an eigenstate~$\Ket{\mathbf{m}}{b}$,
\begin{equation}
\bar{\mathcal{T}}\Ket{\mathbf{m}}{b} = x^{(\mathbf{m})}\Ket{\mathbf{m}}{b}\,.
\end{equation}
Extracting the relevant physics in terms of the physical~$a$ particles may be accomplished by transforming back to~$\{a_{\mathbf{k}}^\dagger,a_{\mathbf{k}}\}$ via the Bogoliubov transformation, Eq.~(\ref{eq:bogoliubov}).

\section{Dominant mode contributions to fluctuations of different magnitudes}
\label{sec:theory}
A useful class of operators to consider is the normal ordered squares of time derivatives of the massless scalar field~$\varphi(t,\mathbf{r})$,
\begin{equation}
\mathcal{T}_{2p'+1}(t,\mathbf{r})=\normord{\tau^{2p'+2}\left[\frac{\partial^{p'}}{\partial t^{p'}}\varphi(t,\mathbf{r})\right]^2}\,. 	
\label{eq:opdef}
\end{equation}
Here $\tau$ is a sampling timescale discussed below.
The factor of~$\tau^{2p'+2}$ ensures~$\mathcal{T}_{2p'+1}(t,\mathbf{r})$ remains dimensionless.  We introduce a more convenient label~$p$, related to the number of time derivatives~$p'$ via
\begin{equation}
p=2p'+1\,.
\end{equation}
Thus,~$p=3$ corresponds to~$\mathcal{T}_3(t,\mathbf{r})=\normord{\tau^4\dot\varphi^2(t,\mathbf{r})}$,~$p=5$ corresponds to~$\mathcal{T}_5(t,\mathbf{r})=\normord{\tau^6\ddot\varphi^2(t,\mathbf{r})}$, and so on.  These operators are of interest because they are related to physical quantities that may be probed experimentally.  For example, the~$p=3$ case can be used to infer the behavior of electromagnetic energy density fluctuations~\cite{Fewster:2012ej}, electromagnetic momentum flux fluctuations~\cite{Huang:2016kmx}, and even fluid density fluctuations~\cite{PhysRevResearch.2.032028}.

The high moments of these operators were analyzed in Ref.~\cite{Fewster:2015hga} for time averaging alone and in Ref.~\cite{PhysRevD.101.025006} for spacetime averaging, which we summarize here.  We consider a class of compactly supported sampling functions with Fourier transforms that asymptotically approach
\begin{equation}
\begin{aligned}
    \hat{f}(\omega)&\sim C_f e^{-\beta|\omega\tau|^\alpha}\,, ~ |\omega\tau|\gg1\,,\\
    \hat{g}(\mathbf{k})&\sim \frac{C_g}{k^{2-\lambda}}e^{-\eta|\mathbf{k}\ell|^\lambda}\,, \; |\mathbf{k}\ell|\gg 1\,.
    \label{eq:ft}
\end{aligned}    
\end{equation}
The constants are assumed to be constrained by~$C_f, C_g, \beta, \eta>0$,~$0<\lambda\leq\alpha<1$, and~$\ell\eta^{1/\lambda}<\tau\beta^{1/\alpha}$.  Here~$\tau$ and~$\ell$ are related to how quickly the sampling functions switch on and off near the bounds of their compact supports.  In this paper,~$\tau$ and~$\ell$ are on the order of the sampling times and lengths, respectively, and may be viewed as such, though this need not be true in general.  The factor of~$k^{\lambda-2}$ in the expression for~$\hat{g}(\mathbf{k})$ is not necessary but is present in our numerical constructions later. 

For~$\mathcal{T}_p(t,\mathbf{r})$ as defined in Eq.~(\ref{eq:opdef}), after spacetime averaging in rectangular coordinates, we find the form in Eq.~(\ref{eq:operator}) with
\begin{equation}
\begin{aligned}
    \bar{F}_{\mathbf{k}\mathbf{k}'} &=\tau^{p+1} \frac{(\omega\omega')^{(p-2)/2}}{V} \hat{f}(\omega-\omega') \hat{g}(\mathbf{k}-\mathbf{k}')\,,\\
    \bar{G}_{\mathbf{k}\mathbf{k}'} &=\tau^{p+1}\frac{(\omega\omega')^{(p-2)/2}}{V} \hat{f}(\omega+\omega') \hat{g}(\mathbf{k}+\mathbf{k}') \,.
\end{aligned}  
\label{eq:fgrectangular}  
\end{equation}
Let us define the $d$th moment of the spacetime-averaged~$\bar{\mathcal{T}}_p$ as
\begin{equation}
\mu_d = \Bra{\mathbf{0}}{a}(\bar{\mathcal{T}}_p)^d \Ket{\mathbf{0}}{a}\,.
\end{equation} 
One may show that
\begin{equation}
\mu_d \approx
\begin{cases}
C_1\int_0^\infty d\omega\, \hat{f}^2(\omega)\omega^{dp+1} &\text{worldline limit } (d\lesssim d_*)\,,\\
C_2\int d^3q \,q^{d(p-2)} \int d^3k \, k \hat{f}^2(q+k) \hat{g}^2(\mathbf{q}+\mathbf{k}) &\text{spacetime-averaged limit }(d\gtrsim d_*)\,.
\end{cases}
\label{eq:momentfreq}
\end{equation}
Here~$C_1$ and~$C_2$ are constants that depend on~$p$ and the sampling functions, and~$k=|\mathbf{k}|$.  Note that the two limits are distinguished by the absence or presence of the effects of space averaging. The lower moments are independent of the space averaging, but the higher moments depend upon  $\hat{g}(\mathbf{k})$. The effect of space averaging is to reduce the rate of growth of the moment as $d$ increases.  The value $d_*$ depends on the ratio of~$\tau\beta^{1/\alpha}$ to~$\ell\eta^{1/\lambda} $ in some power law behavior: the smaller this ratio, the sooner the effects of space averaging emerge and the earlier the transition to the spacetime-averaged limit.  We may intuitively understand this behavior by noting that in the limit of no space averaging,~$\ell\to0$, we reduce to~$d_*\to\infty$.  The expression in the worldline limit is derived in Ref.~\cite{Fewster:2015hga} while the expression in the spacetime-averaged limit can be easily generalized from the~$p=3$ case in Ref.~\cite{PhysRevD.101.025006} by noting from Eq.~(\ref{eq:fgrectangular}) that higher~$p$ merely includes more factors of angular frequency.

These moments may be related to those of a probability distribution~$P(x)$,
\begin{equation}
\mu_d = \int_{-\infty}^\infty dx \, x^d P(x)\,
\label{eq:momentx}
\end{equation}
from which we may infer~\cite{Fewster:2015hga,PhysRevD.101.025006}
\begin{equation}
P(x)\sim c_0 x^b e^{-ax^c}\,,~x\gg1\,,
\label{eq:prob}
\end{equation}
where~$c=\alpha/p$ in the worldline limit and~$c=\alpha/(p-2)$ in the spacetime-averaged limit. We will not be concerned with the remaining constants, which are predicted in the worldline limit but are less clear in the spacetime-averaged limit.  Similarly to Eq.~(\ref{eq:momentfreq}), the transition from the worldline limit to the spacetime-averaged limit occurs around~$x\approx x_*$, depending again on the ratio of~$\tau\beta^{1/\alpha}$ to~$\ell\eta^{1/\lambda}$ and may be generalized from Ref.~\cite{PhysRevD.101.025006} to be
\begin{equation}
    x_*\approx\left(\frac{\tau\beta^{1/\alpha}}{\ell\eta^{1/\lambda}}\right)^p\,.
    \label{eq:xstar}
\end{equation}
We may now see that the transition from worldline to spacetime-averaged  behavior is linked both to the ratio $\tau/\ell$ and to the relative rates of decay of $\hat{f}(\omega)$ and of 
$\hat{g}(\mathbf{k})$ functions of their arguments. If $\lambda \leq \alpha$, and $\beta \approx \eta$, then we see from Eq.~(\ref{eq:ft}) that  $\hat{f}(\omega)$ decreases more rapidly with 
increasing $\omega$ than does $\hat{g}(\mathbf{k})$ with increasing $|\mathbf{k}|$ when $\tau > \ell$.

We now wish to estimate the dominant frequency contributions to~$\mu_d$ from Eq.~(\ref{eq:momentfreq}) and the dominant eigenvalue contributions from Eq.~(\ref{eq:momentx}) by finding the peaks of the integrands in these equations.  Assuming that~$d\gg1$, the integrals in Eq.~(\ref{eq:momentfreq}) may be approximated as
\begin{equation}
\mu_d \approx
\begin{cases}
C_1C_f^2\int_0^\infty d\omega\, e^{-2\beta(\omega\tau)^\alpha}\omega^{dp+1} &\text{worldline limit } (d\lesssim d_*)\,,\\
6\pi^2\alpha^{-4}C_2C_f^2C_g^2\int_0^\infty dk\, e^{-2\beta(k\tau)^\alpha} k^{d(p-2)}  &\text{spacetime-averaged limit }(d\gtrsim d_*)\,.
\end{cases}
\end{equation}
Notice that~$\lambda$ and~$\eta$ do not explicitly appear in the spacetime-averaged limit here despite the presence of~$\hat{g}^2(\mathbf{q}+\mathbf{k})$ in Eq.~(\ref{eq:momentfreq}).  When~$d\gg1$, the integrals may be approximated by taking~$k\gg1$, and in this limit we have~$k^\alpha\gg k^\lambda$ for~$\alpha > \lambda$.  We may now estimate~$\omega_d$, the dominant frequency contribution to the~$d$th moment, as the frequency that maximizes the integrands, giving
\begin{equation}
\omega_d \approx 
\begin{cases}
\left(\frac{dp}{2\beta\alpha\tau^\alpha}\right)^{1/\alpha} &\text{worldline limit,}\\
\left(\frac{d(p-2)}{2\beta\alpha\tau^\alpha}\right)^{1/\alpha} &\text{spacetime-averaged limit.}
\end{cases}
\label{eq:omegad}
\end{equation}
Similarly, we may estimate~$x_d$, the dominant eigenvalue contribution to the~$d$th moment, as the eigenvalue that maximizes the integrand in Eq.~(\ref{eq:momentx}), finding
\begin{equation}
x_d \approx \left(\frac{d}{ac}\right)^{1/c}\,.
\label{eq:xd}
\end{equation}
Here we recall that in the worldline limit,~$c=\alpha/p$, whereas in the spacetime-averaged limit,~$c=\alpha/(p-2)$.  We combine Eqs.~(\ref{eq:omegad}) and~(\ref{eq:xd}) to find
\begin{equation}
\begin{aligned}
\omega_d \propto
  \begin{cases}
    x_d^{1/p} &\text{worldline limit}\,,\\
    x_d^{1/(p-2)} & \text{spacetime-averaged limit}\,.
  \end{cases}
  \end{aligned}
  \label{eq:theory}
\end{equation}
As the value of~$a$ is not well predicted, we are not interested in the proportionality constant, which, at any rate, may be inferred numerically.  The crucial behavior here is the power law behavior.  The transition from the worldline to spacetime-averaged limit is expected to be given by Eq.~(\ref{eq:xstar}), though we do not have enough data to verify this prediction.  Reference~\cite{PhysRevD.103.125014} suggests the exponent in Eq.~(\ref{eq:xstar}) is well predicted but the proportionality constant is not, an observation that likely holds in our case.

Note that the analytical estimates in this section do not yet have a clear physical interpretation.  Equation~(\ref{eq:theory}) relates the dominant frequency and eigenvalue contributions, but these dominant contributions have not yet been shown to have meaningful interpretations.  One straightforward interpretation of~$x_d$ is to consider it as an estimate of the magnitudes of large fluctuations,~$x$.  However, a similarly straightforward interpretation of~$\omega_d$ is less clear, as we do not yet have a clear understanding of the frequencies contributing to different fluctuations, the topic of the next section.  We will find that~$\omega_d$ gives an accurate estimate of the maximum frequency contributing to a fluctuation of magnitude~$x\approx x_d$, rather than other possibilities such as the most likely frequency to contribute to fluctuations of that magnitude.

\section{Numerical estimates of mode contributions to large fluctuations}
\label{sec:results}
In this section we are primarily concerned with the characterization of the frequency modes that contribute to large vacuum fluctuations.  We will find for any given quantum fluctuation, a wide range of frequency modes contribute, necessitating the introduction of some measure that encodes this information.  In doing so, we will show that Eq.~(\ref{eq:theory}) predicts the greatest angular frequency~$\omega_d$ that substantially contributes to a vacuum fluctuation of magnitude~$x_d$.

In the numerical simulations, we work in spherical coordinates to take advantage of the assumed spherical symmetry of the spatial sampling function~$g(\mathbf{r})$.  The spherical coordinate equivalent of Eq.~(\ref{eq:fgrectangular}) is~\cite{PhysRevD.103.125014}
\begin{equation}
\begin{aligned}
\bar{F}_{nlm,n'l'm'}&=\frac{\tau^{p+1}(\omega_{nl}\omega_{n'l})^{(p-2)/2}\,\delta_{l,l'}\, \delta_{m,m'}}{4\pi R^3j_{l+1}(\omega_{n'l}R)j_{l+1}(\omega_{nl}R)}\,\hat{f}\left(\omega_{nl}-\omega_{n'l}\right)\int_{-1}^1 dy\, \hat{g}\left(\sqrt{\omega_{nl}^2 + \omega_{n'l}^2 -2\omega_{nl}\omega_{n'l} y}\right) P_l(y)\,,\\
\bar{G}_{nlm,n'l'm'}&=\frac{-\tau^{p+1}(\omega_{nl}\omega_{n'l})^{(p-2)/2}\,\delta_{l,l'}\, \delta_{m,-m'}}{4\pi R^3j_{l+1}(\omega_{n'l}R)j_{l+1}(\omega_{nl}R)}\,\hat{f}\left(\omega_{nl}+\omega_{n'l}\right)\int_{-1}^1 dy\, \hat{g}\left(\sqrt{\omega_{nl}^2 + \omega_{n'l}^2 -2\omega_{nl}\omega_{n'l} y}\right) P_l(y)\,.
\end{aligned}
\end{equation}
Here~$n$ is a positive integer,~$l$ is a non-negative integer, and~$m$ is an integer satisfying~$|m|\leq l$.  The function~$j_l(kr)$ is the~$l$th spherical Bessel function, the function~$P_l(y)$ is the~$l$th Legendre polynomial,~$\delta_{l,l'}$ is the Kronecker delta,~$R$ is the radius of the bounding sphere of the quantized system, and
\begin{equation}
\omega_{nl}= \frac{z_{nl}}{R}\,,
\end{equation}
where~$z_{nl}$ is the~$n$th zero of~$j_l(kr)$.  Reference~\cite{PhysRevD.103.125014} suggests reliable results may be obtained by setting~$l=m=0$, giving the simpler form
\begin{equation}
\begin{aligned}
\bar{F}_{nn'}&=\frac{ \tau^{p+1} \pi^{p-3}(nn')^{(p-2)/2}}{4(-1)^{n+n'}R^{p-1}} \hat{f}\left(\frac{(n- n')\pi}{R}\right) \int_{|n-n'|\pi/R}^{(n+n')\pi/R} dk\,k \hat{g}(k)\,,\\
\bar{G}_{nn'}&=\frac{-\tau^{p+1} \pi^{p-3}(nn')^{(p-2)/2}}{4(-1)^{n+n'}R^{p-1}} \hat{f}\left(\frac{(n+n')\pi}{R}\right) \int_{|n-n'|\pi/R}^{(n+n')\pi/R} dk\,k \hat{g}(k)\,.
\end{aligned}
\end{equation}
Functions~$\hat{f}(\omega)$ and~$\hat{g}(k)$ with the behavior in Eq.~(\ref{eq:ft}) may be constructed numerically~\cite{Fewster:2015hga,PhysRevD.101.025006}.  A particular implementation is detailed in Sec. IVA1 and IVA2 of Ref.~\cite{PhysRevD.103.125014}, which we continue using here.  A choice of~$\tau$,~$\alpha$, and the sampling time specifies~$\hat{f}(\omega)$ while a choice of~$\ell$,~$\lambda$, and the sampling length specifies~$\hat{g}(r)$.    We consider the specific case~$\alpha=\lambda=0.5$ in Eq.~(\ref{eq:ft}), which is less susceptible to numerical error.  As in Ref.~\cite{PhysRevD.103.125014}, we work in~$\tau=1$ units, and in these units the sampling time and radius are~$2$ and~$0.0028$, respectively, while the radius of the bounding sphere,~$R$, is 0.88.  In our construction, the parameter~$\ell$ is equal to the radius of the sampling volume, i.e.,~$\ell=0.0028$.  Note that these parameter choices are suboptimal given the assumption of flat space: the boundary of the quantization sphere is detectable by an observer at the origin, as the light travel time to the boundary and back is shorter than the sampling time.  Numerical instabilities prevent a more optimal parameter selection, a topic for future investigation.

\subsection{Frequency spectra of $a$ particles in eigenstates of the operator}
\begin{figure}[tb]
    \centering
    \includegraphics[width=0.45\linewidth]{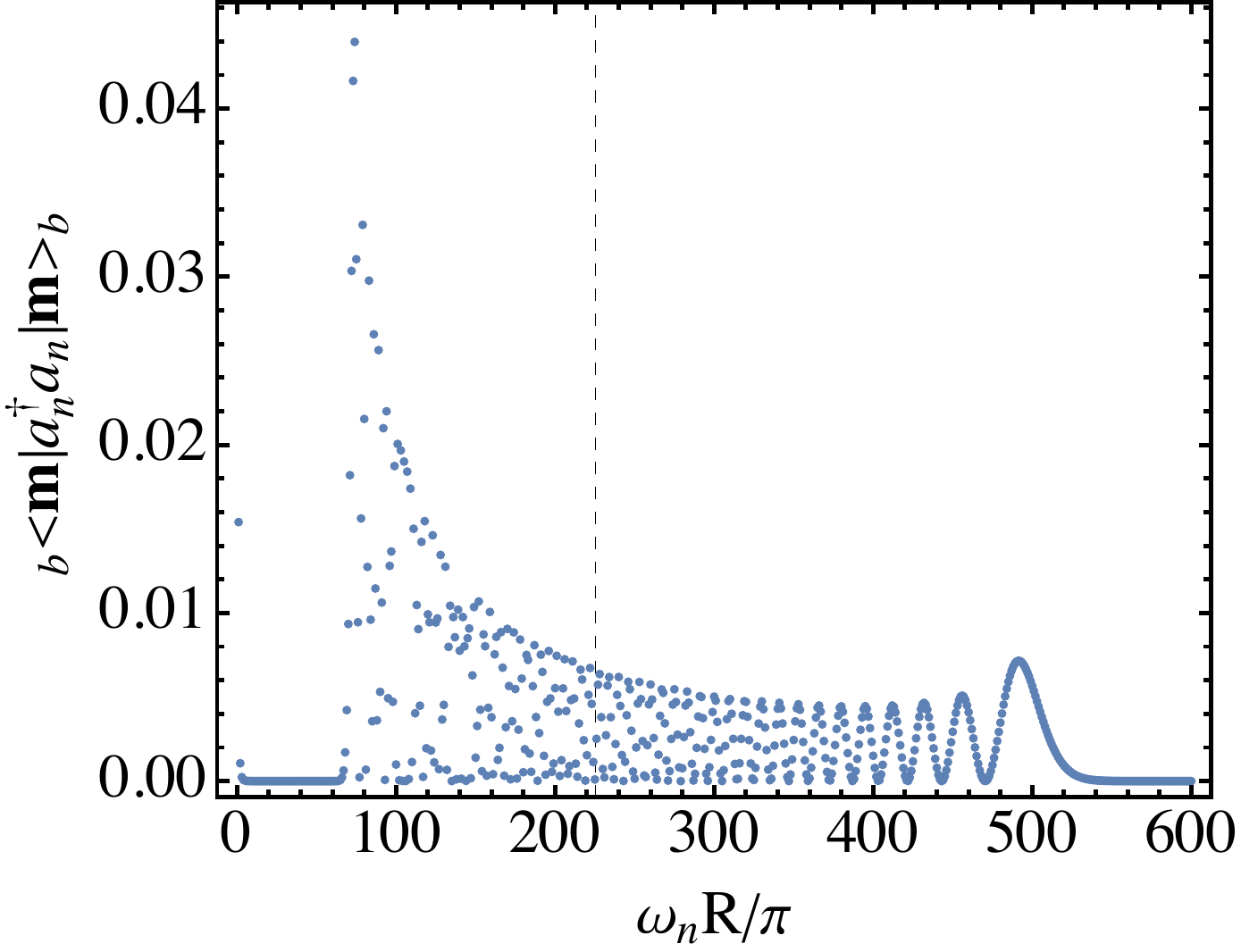}
        \caption{Plotted is a frequency spectrum for an eigenstate of the~$p=5$ operator with~$n=1,2,\ldots,600$ in the~$b$ two-particle sector.  The vertical dashed lined denotes the location of~$\omega_c$ defined in Eq.~(\ref{eq:omegac}).  The observed behavior does not change drastically with the parameter choices nor with higher~$b$ particle sectors.  The shapes of the spectra are roughly consistent for different eigenstates, qualitatively differing by translations along the horizontal axis, an effect that will be further discussed in Sec.~\ref{sec:omegac}.}
            \label{fig:spectra}
\end{figure}

Recall that the physical particles of interest are the~$a$ particles associated with the Minkowski vacuum state~$\Ket{\mathbf{0}}{a}$, but the magnitudes of the quantum fluctuations,~$x$, are more easily expressed in the eigenstates given by the~$b$ particle number states~$\Ket{\mathbf{m}}{b}$.  We would like to construct the frequency spectra of~$a$ particles in the eigenstates~$\Ket{\mathbf{m}}{b}$ to better understand the frequencies of~$a$ particles that contribute to various quantum fluctuations.  To do so, let us consider the mean number of~$a$ particles with frequency~$\omega=k$ in some eigenstate~$\Ket{\mathbf{m}}{b}$.  We may calculate this quantity by evaluating
\begin{equation}
\Bra{\mathbf{m}}{b}a^\dagger_{\mathbf{k}}a_{\mathbf{k}}\Ket{\mathbf{m}}{b} = \sum_{\mathbf{k'}}[A^2_{\mathbf{k}\mathbf{k'}}m_{\mathbf{k'}} + B^2_{\mathbf{k}\mathbf{k'}}(m_{\mathbf{k'}}+1)]\,,
\label{eq:meanfreqa}
\end{equation}
where the Bogoliubov transformation, Eq.~(\ref{eq:bogoliubov}), has been used.  Here~$m_{\mathbf{k}}$ labels the particle content of the eigenstates~$\Ket{\mathbf{m}}{b}$ with the convention shown in Eq.~(\ref{eq:bket}).  This result describes the relation between the~$b$ particle content, which is encoded in the eigenstates~$\Ket{\mathbf{m}}{b}$, and the~$a$ particle content, which are the physically interesting particles.  Figure~\ref{fig:spectra} shows an example of an eigenstate with two~$b$ particles, i.e.,~$\sum_{\mathbf{k}} m_{\mathbf{k}}=2$.  There is some indication that the total mean number of~$a$ particles may be equal to the total number of~$b$ particles, but more analysis is needed to confirm this.  The frequency spectrum displays complex structure that eludes straightforward interpretation, though the higher frequency modes have diminishing contributions, a behavior that is generic to the spectra that we will investigate in more detail.  While the mean numbers of~$a$ particles fall sharply to negligible values at small and high frequencies, we cannot rule out limitations in our simulations, so we leave further discussion of the consequences for future work.  Eigenstates with more than two~$b$ particles do not display different behavior, so we will consider only the two-particle sector in our simulations.  The higher particle sectors are generally less probable, as argued in Ref.~\cite{PhysRevD.103.125014}, and should not significantly affect our conclusions.  

\subsection{Characteristic frequencies of frequency spectra}
\label{sec:omegac}
Because the frequency spectra are too complex to analyze directly, we would like to extract a single characteristic frequency from each spectrum.  Here we proceed with the simplest characterization,
\begin{equation}
\omega_c = \frac{\sum_{\mathbf{k}} \left(\Bra{\mathbf{m}}{b} a^\dagger_\mathbf{k}a_\mathbf{k}\Ket{\mathbf{m}}{b}\,\omega_\mathbf{k}\right)}{\sum_{\mathbf{k}} \,\Bra{\mathbf{m}}{b}  a^\dagger_\mathbf{k}a_\mathbf{k}\Ket{\mathbf{m}}{b}}\,,
\end{equation}
which is merely an average of all frequencies, weighted by the mean number of~$a$ particles at different frequencies.  The denominator normalizes the total mean number to unity.  For our case with $l=m=0$ and~$n=1\sim N$, we get
\begin{equation}
\omega_c = \frac{\pi}{R}\frac{\sum_{n=1}^N \left(\Bra{\mathbf{m}}{b} a^\dagger_n a_n \Ket{\mathbf{m}}{b}\,n\right)}{\sum_{n=1}^N \,\Bra{\mathbf{m}}{b}  a^\dagger_n a_n\Ket{\mathbf{m}}{b}}\,.
\label{eq:omegac}
\end{equation}

\begin{figure}[hp]
    \centering
    \begin{minipage}[c]{\linewidth}
    \includegraphics[width=0.45\linewidth]{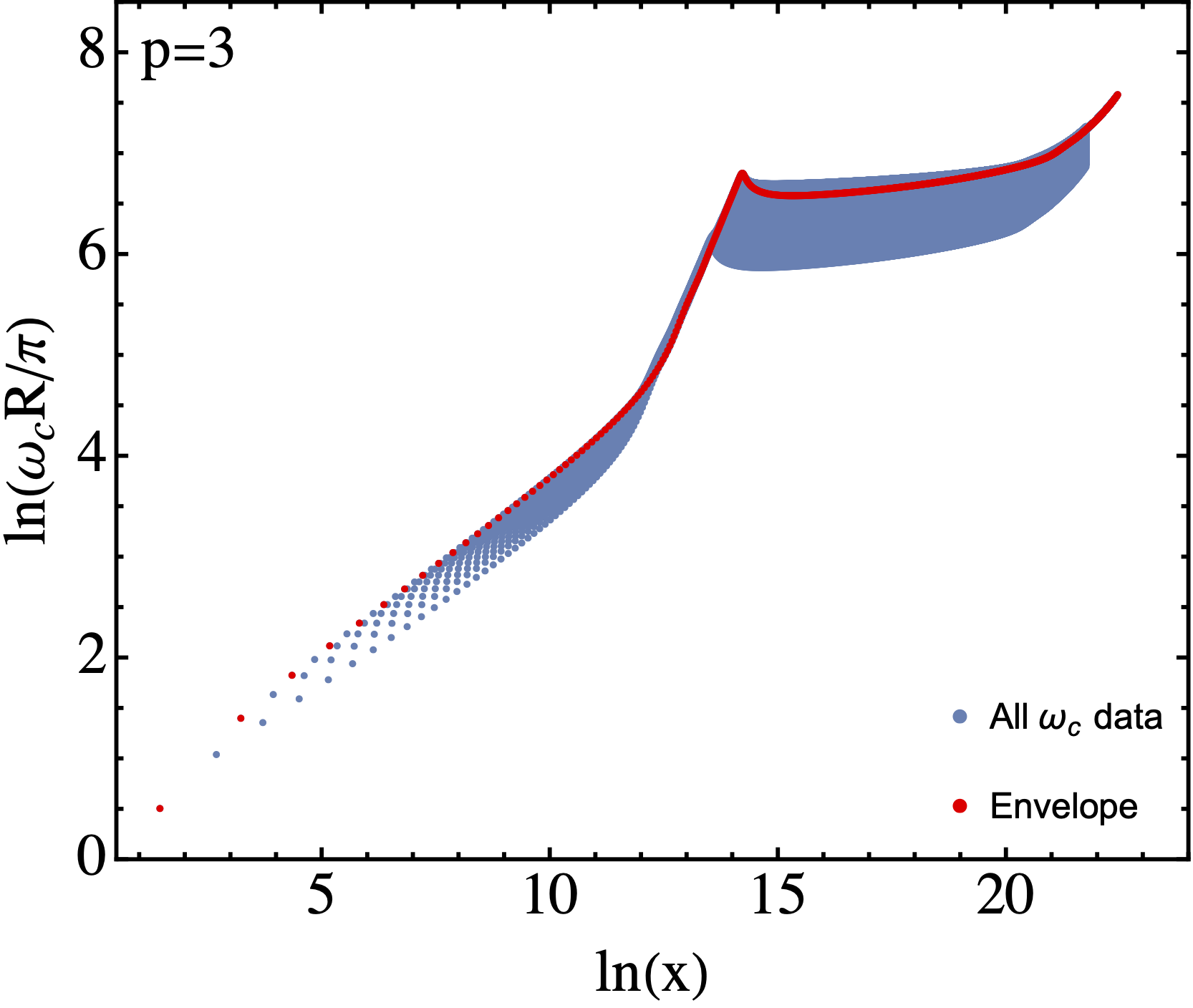}
    \end{minipage}
    \\
    \vspace{2mm}
    \begin{minipage}[c]{\linewidth}
    \includegraphics[width=0.45\linewidth]{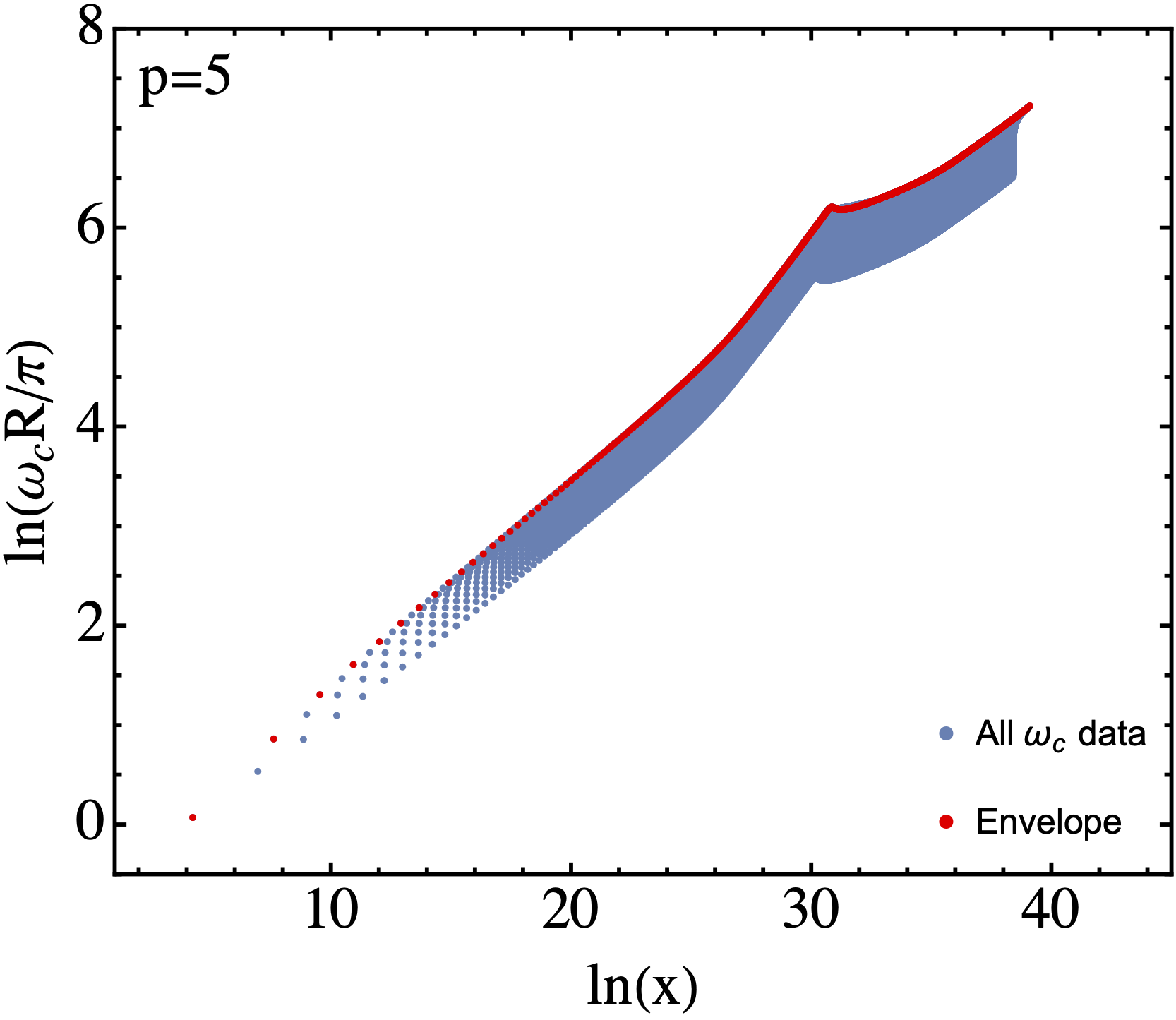}
    \end{minipage}
    \\    
    \vspace{2mm}
    \begin{minipage}[c]{\linewidth}
    \includegraphics[width=0.45\linewidth]{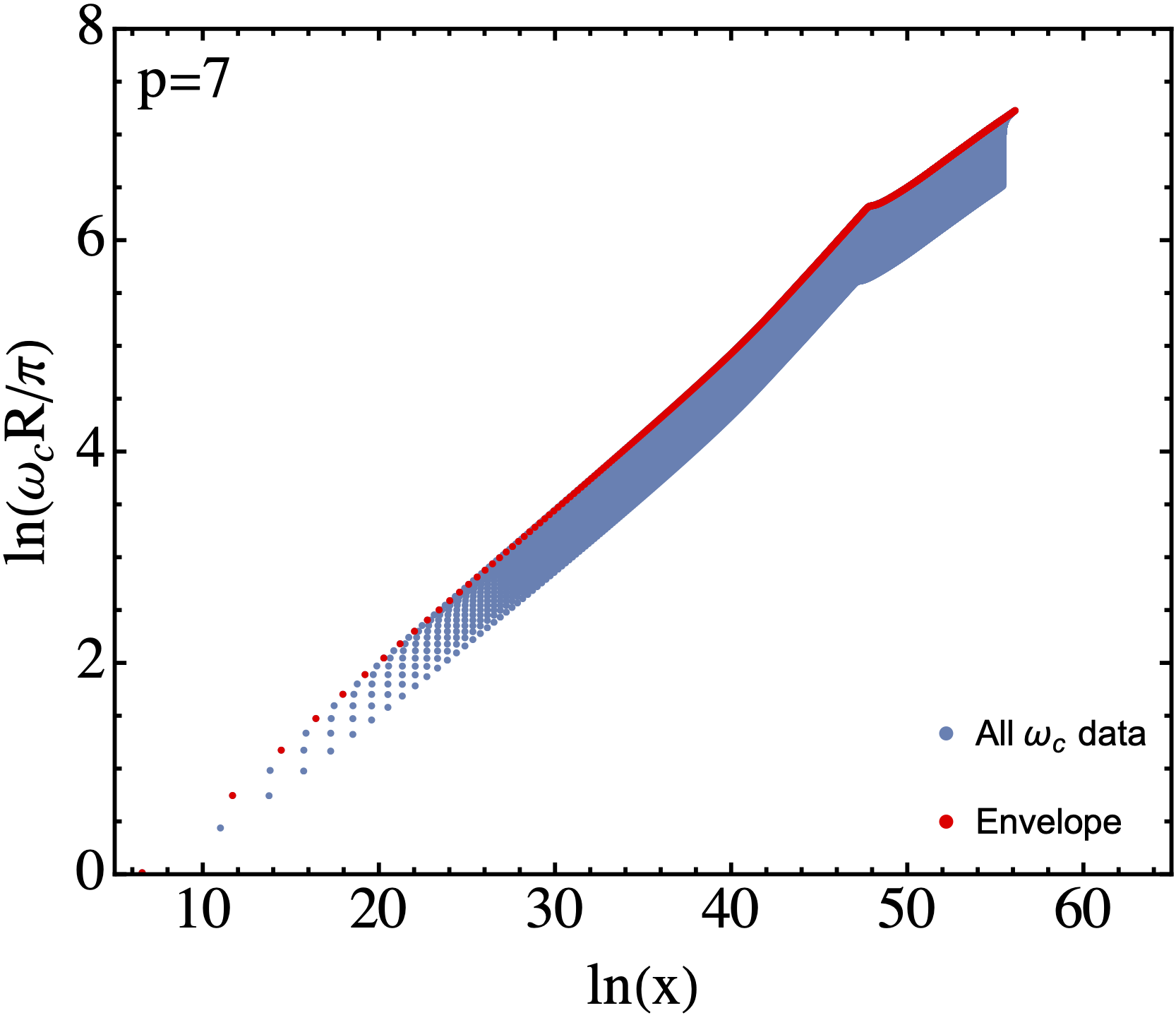}
        \caption{Plotted are the data sets for, from the top to bottom panel,~$p=3$, 5, and 7, respectively, that will be used for detailed numerical analysis later in Sec.~\ref{sec:fit}.    Highlighted in red is the subset of eigenstates where only one frequency mode has nonzero particle content.  The data shown are strictly in the~$b$ two-particle sector.}
            \label{fig:data}
    \end{minipage}
\end{figure}

We will find it convenient to look at the frequencies scaled by~$R/\pi$ because~$1\leq\omega_cR/\pi\leq N$, where~$N$ is the total number of modes in the simulation.  In Fig.~\ref{fig:data}, we show plots of the characteristic frequencies against the eigenvalues on log-log scales.  Observe that there is clear structure to the data sets, suggesting the weighted average~$\omega_c$ encodes sufficient information to draw broad conclusions about the frequency content of quantum fluctuations.  For any given fluctuation of magnitude roughly~$x$, we find a wide range of frequencies that contribute to that fluctuation.  Interestingly, except for~$p=3$, the envelopes of the data sets are given by a specific subset of eigenstates, those where both~$b$ particles are in the same frequency mode.  The anomalous behavior of~$p=3$ remains under investigation, and it remains unclear why there is a region ($12 < \ln x < 13$ in the top panel of Fig.~\ref{fig:data})  where the envelope is not given by this subset of states.  Regardless, in this region the upper and lower bounds on the characteristic frequencies almost coincide, so our numerical results should still suffice for rough estimates.

However, given the complex behavior of the spectra, we may be concerned whether a finite-mode computation can be used in physical applications, which are better described by the~$N\to \infty$ limit.  We first recall our observation from Fig.~\ref{fig:spectra}: generically, higher frequency modes have diminishing contributions.  We supplement this observation with a simple argument.  As higher frequency modes are included, the characteristic frequency~$\omega_c$, a weighted average of all frequencies, increases.  If~$\omega_c$ is not bounded above, then in the~$N\to\infty$ limit, we have~$\omega_c\to\infty$ for all quantum fluctuations, a nonsensical result.  We thus expect that for any given fluctuation, only some frequency modes have substantial contributions. 

\begin{figure}[tb]
    \centering
    \begin{minipage}[c]{\linewidth}
    \includegraphics[width=0.45\linewidth]{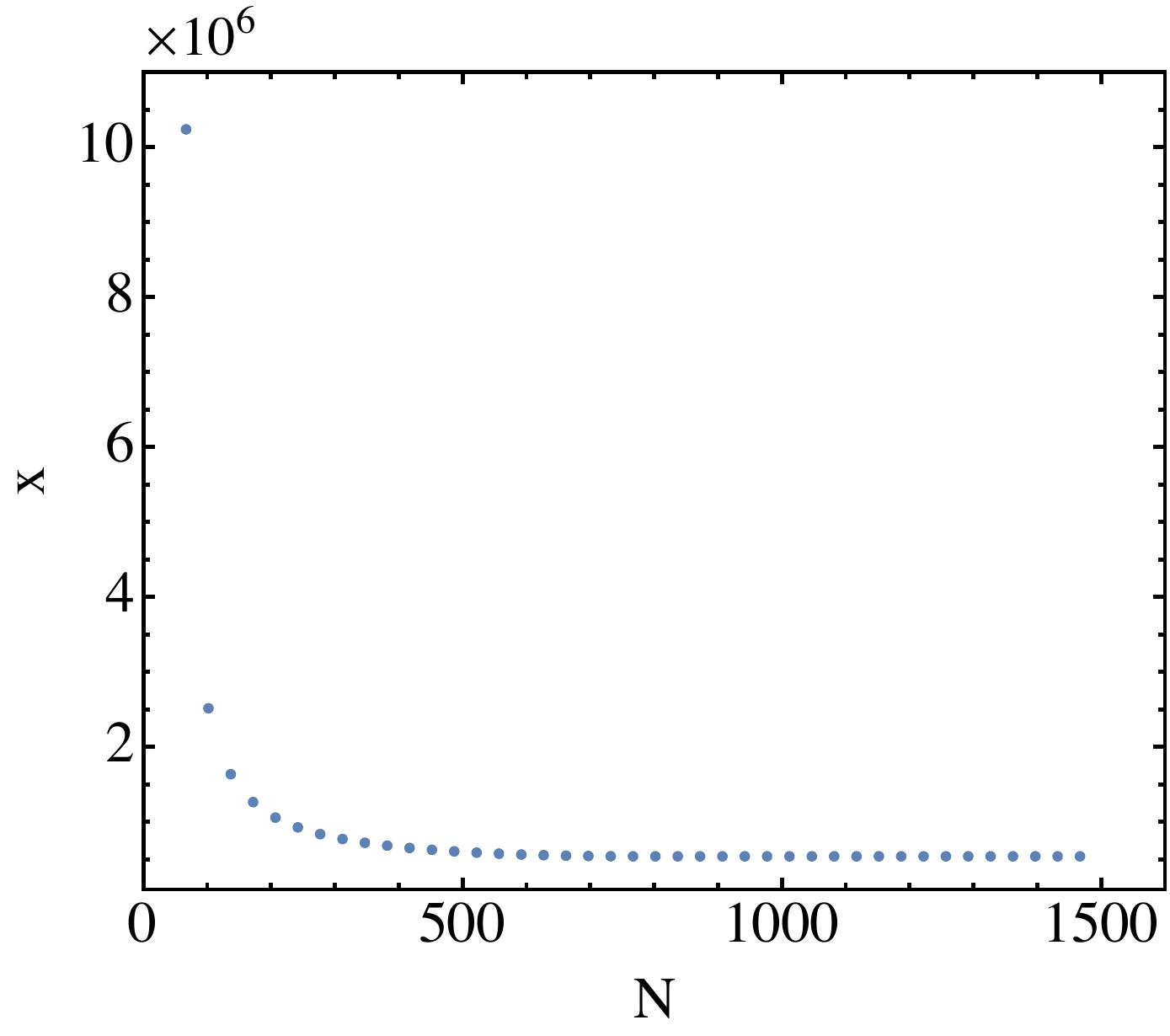}
    \end{minipage}
    \\
    \vspace{2mm}
    \begin{minipage}[c]{\linewidth}
    \includegraphics[width=0.45\linewidth]{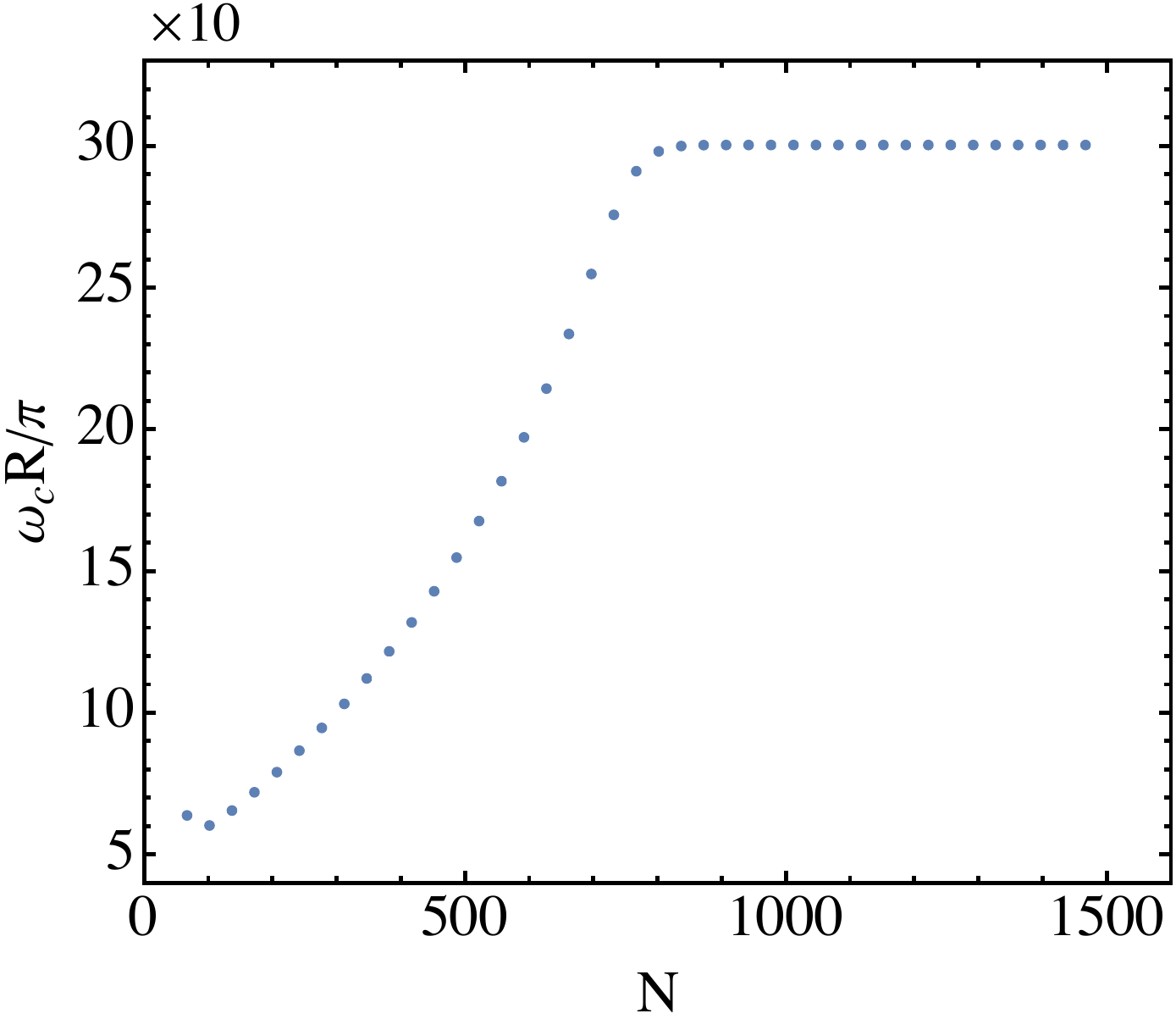}
        \caption{Plotted are the eigenvalues and characteristic frequencies as a function of the total number of modes~$N$, where the mode selection is~$n=1\sim N$,~$l=m=0$, for the case~$p=3$ and~$\ell=0.0028$, identical to the choices later in Sec.~\ref{sec:results}.  At sufficiently high~$N$, both~$x$ and~$\omega_c$ converge, a behavior that generically holds for arbitrary parameters and eigenstates.}
            \label{fig:converge}
    \end{minipage}
\end{figure}

We may also draw this conclusion numerically in the following manner.  The aforementioned argument suggests that as~$N$ increases,~$\omega_c$ will converge to some stable value.  We confirm this behavior in Fig.~\ref{fig:converge} for both~$\omega_c$ and~$x$.  These results suggest that our numerical analysis of finite-mode systems can be applied to any physical system of interest, provided we limit our analysis to the region that has converged.  Recalling that we can identify the states that give the envelope, we focus our analysis on the envelope of the data sets.  In doing so, we will find in Sec.~\ref{sec:fit} that~$\omega_d$ in Eq.~(\ref{eq:theory}) may be interpreted as setting an upper bound on the characteristic frequencies~$\omega_c$.  Note that the end of the converged region may be visually distinguished by the kink in the slopes at large~$x$ in Fig.~\ref{fig:data}.  

At this point, it may be worth pointing out some subtleties to our interpretation of the theoretical prediction in Eq.~(\ref{eq:theory}).  Note that~$\omega_c$ is a weighted average, and thus frequencies greater than~$\omega_c$ still contribute, so it is more precise to say that~$\omega_d$ sets upper bounds on the frequencies that contribute substantially, with the understanding that higher frequency modes may have smaller, and eventually negligible, contributions.  We should also note that for any fluctuation of magnitude~$x$, there appears to be exactly one corresponding eigenstate and thus one or two corresponding $b$-mode frequencies, in the two-particle sector. 
However, there can be a wide range of $a$-mode frequencies, which is the focus of our discussion. Thus,  Eq.~(\ref{eq:theory}) predicts upper bounds on the $a$-mode frequencies that contribute significantly to fluctuations with magnitudes of order $x$.

\subsection{Numerical fits to the envelope}
\label{sec:fit}
Let us now analyze the numerical data to confirm the theoretical predictions, Eq.~(\ref{eq:theory}).  As we are primarily concerned with the power law behavior, we take natural logarithms to find
\begin{equation}
\ln(\omega_dR/\pi) \approx
\begin{cases}
\frac{1}{p}\ln(x_d)&\text{worldline limit,}\\
\frac{1}{p-2}\ln(x_d) &\text{spacetime-averaged limit.}
\end{cases}
\label{eq:logtheory}
\end{equation}
Note that we have neglected a nonzero additive constant that emerges from the proportionality constant in Eq.~(\ref{eq:theory}) and our choice to rescale by a factor of~$R/\pi$.  An adequate data set must contain sufficient data in both the worldline and spacetime-averaged limits to perform numerical fits.  Given computational constraints, we must make a compromise to get enough data in both regions.  Preliminary analysis following the method in Sec.~\ref{sec:omegac} suggests that the convergence rates as functions of~$N$ increase as~$p$ increases.  Consequently, for similar amounts of data points, we must go to higher~$N$ for smaller values of~$p$.  We further recall that the transition from the worldline to spacetime-averaged limits scales as~$\tau/\ell$ raised to some power, so smaller values of~$\ell$ give us more data in the worldline limit.  These two observations underlie our parameter choices for our data sets.

\begin{figure}[htbp]
    \centering
    \begin{minipage}[c]{\linewidth}
    \includegraphics[width=0.45\linewidth]{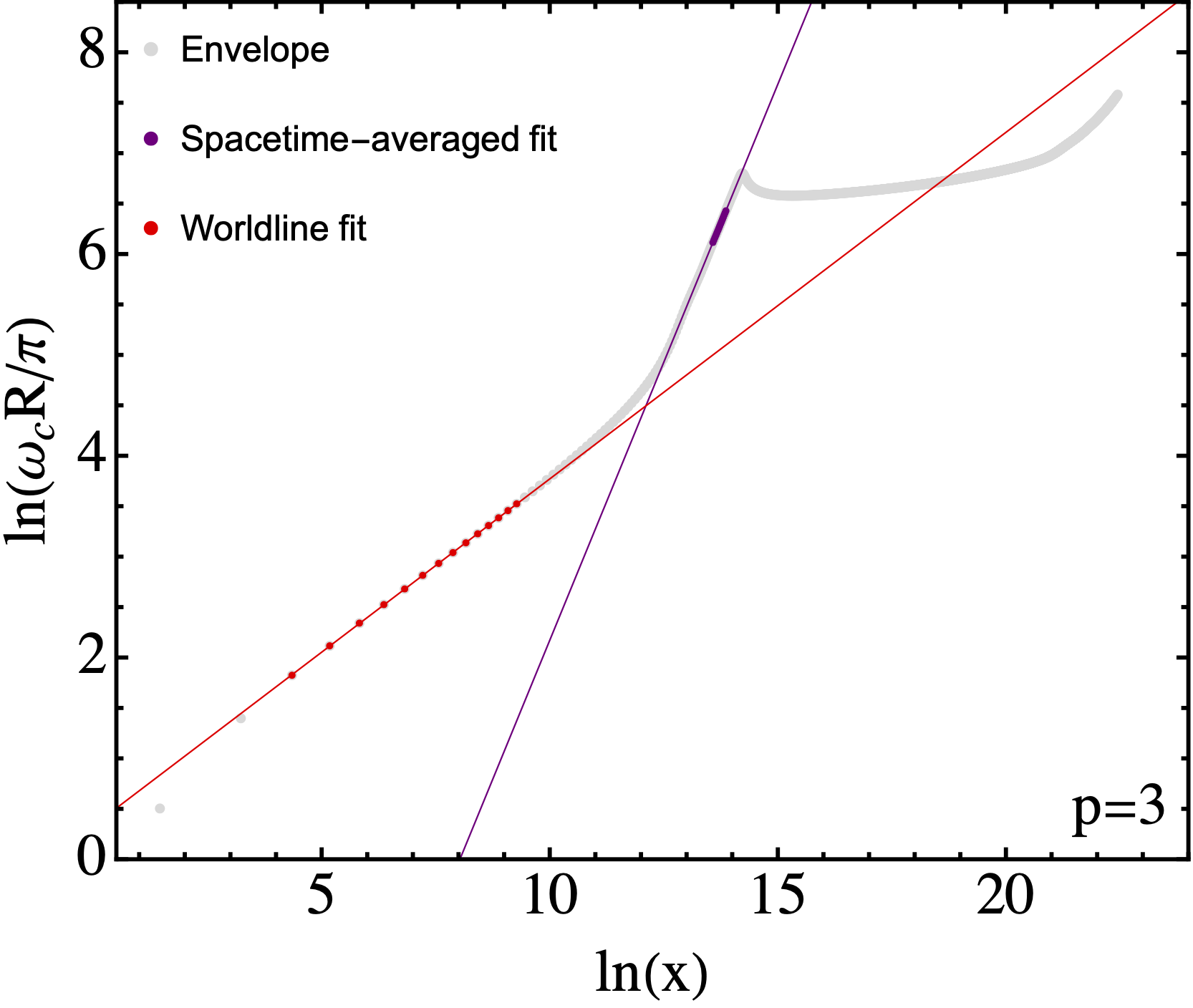}
    \end{minipage}
    \\
    \vspace{2mm}
    \begin{minipage}[c]{\linewidth}
    \includegraphics[width=0.45\linewidth]{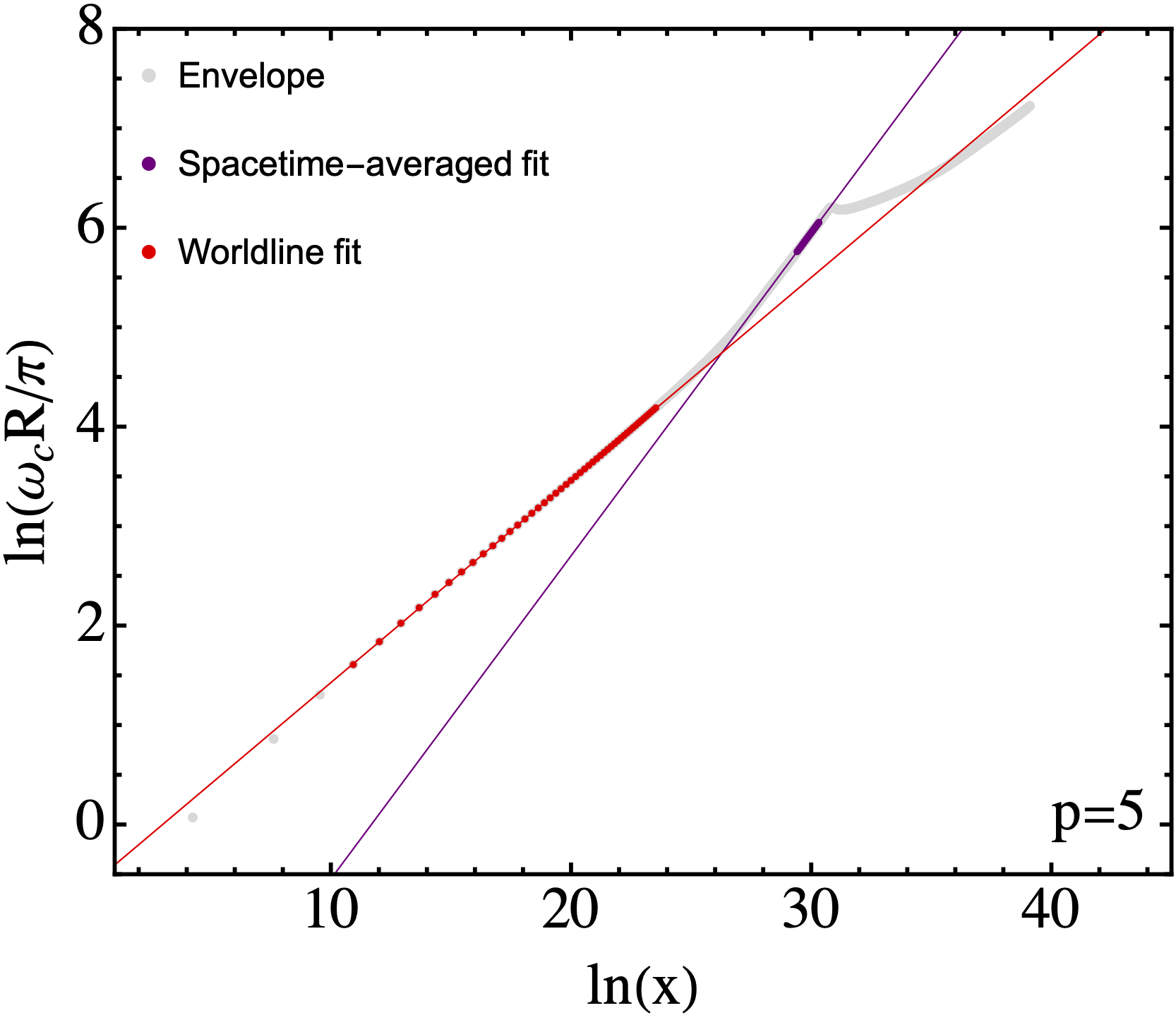}
    \end{minipage}
    \\    
    \vspace{2mm}
    \begin{minipage}[c]{\linewidth}
    \includegraphics[width=0.45\linewidth]{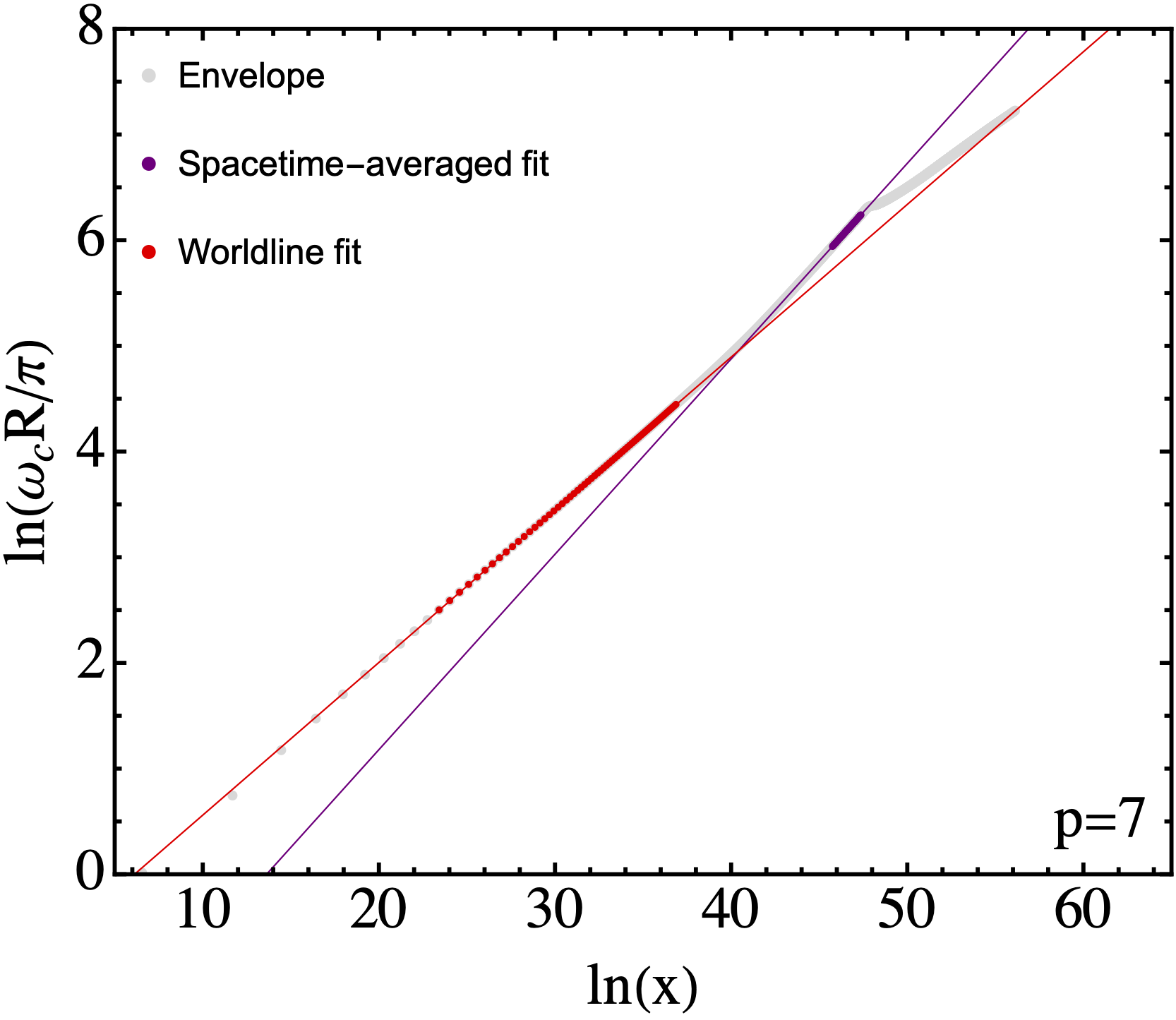}
        \caption{Plotted are the linear fits to the data sets for, from the top to bottom panel,~$p=3$, 5, and 7, respectively.  Here we only show the envelopes in Fig.~\ref{fig:data}.}
            \label{fig:fit}
    \end{minipage}
\end{figure}

We interpret~$\omega_d$ as the envelope of the characteristic frequencies~$\omega_c$ and~$x_d$ as the fluctuation magnitudes~$x$.  For the three data sets shown in Fig.~\ref{fig:data}, we perform least squares linear regressions to the envelopes of the converged regions of the worldline and spacetime-averaged limits, which we may easily determine by eye as the two distinct slopes in each plot.  The transition location, Eq.~(\ref{eq:xstar}), may be analyzed following the method in Sec.~IV B of Ref.~\cite{PhysRevD.103.125014}, which we do not consider here due to the lack of data.  The linear fits are shown in Fig.~\ref{fig:fit}, and the best fit values and statistical errors are compiled in Table~\ref{tab:fit}.  We find fairly good consistency between the predictions and numerical results, verifying our interpretation of Eq.~(\ref{eq:theory}) as setting upper bounds on the frequencies that substantially contribute to fluctuations of varying magnitudes.  Note that full consistency in the spacetime-averaged limit can be obtained by increasing~$\ell$, at the expense of the worldline data.  Full consistency in the worldline limit has not been theoretically explored in much depth, and in particular the~$\ell$ dependence of the onset of the worldline limit is unknown.  Reducing~$\ell$ is thus not guaranteed to provide more data in the worldline limit, and a reliable method to procure better data in this limit is a topic for future work.
As a final comment, note that in choosing to fit strictly to the converged region, our numerical results are not dependent on~$N$ and by extension the higher frequency modes that are omitted from our finite-mode computation.  Our numerical verification of the theoretical prediction, Eq.~(\ref{eq:theory}), is thus mode independent and may be applied to systems with higher degrees of freedom.

\begin{table*}[hbt]
\begin{center}
\caption{Fit results of the least squares linear regressions in the worldline and spacetime-averaged limits.  In all data sets we have~$\ell=0.0028$ whereas~$N$ has been increased for~$p=3$ due to the slower convergence of the eigenvalues and characteristic frequencies.  The number of points in each fit ranges from 14 to 101, with fewer points for smaller~$p$ due to slower convergence.}
\vspace{0.2cm}
\begin{tabular}{cccccc}
\hline
Case&Value of~$N$&Limit&Predicted slope&Fitted slope &Standard error\\          
\hline
$p=3$&2000& Worldline & 0.3333 & 0.3436 & 0.0005\\
      && Spacetime averaged & 1.00 & 1.10 & 0.003\\
$p=5$ & 1400 & Worldline & 0.2000 & 0.2037 & 0.0002\\
      && Spacetime averaged & 0.3333 & 0.3249 & 0.0001\\
$p=7$& 1400&Worldline & 0.14286 & 0.14447 & 0.00007\\
     && Spacetime averaged & 0.20000 &0.18493 & 0.00001\\     
\hline
\label{tab:fit}
\end{tabular}
\end{center}
\end{table*}

\section{Physical applications}
\label{sec:application}
In most models proposing physical effects arising from large vacuum fluctuations, an interaction between a probe and the particles of quantum field is assumed.  The details of these interactions often depend crucially on the frequencies of the quantum particles because the eigenstates of spacetime-averaged quadratic operators are multimode squeezed vacuum states with nonzero particle content, and it is the effects of these particles that are potentially observable.  We have discovered that we may characterize the particle frequencies of fluctuations of various magnitudes, which may aid the development of better experimental proposals. 

Let us consider a brief example how one may take advantage of the results in prior sections.  The case~$p=3$ is particularly interesting due to its broad applicability, and here we explore which photons arising from energy density fluctuations may be trapped in a metal cavity.  Let~$u$ be the energy density of the massless scalar field averaged in some region of spacetime.  The moments of~$u$ are expected to behave similarly to those of~$\overline{\dot\varphi^2}$,\footnote{In fact, for the case of time averaging alone, the relationship between the moments of the energy density and the moments of the square of the time derivative of the massless scalar field is found analytically in Ref.~\cite{Fewster:2012ej}.} so in our notation the dimensionless measures of the fluctuation magnitudes,~$x$, are related to the actual energy density measurements via
\begin{equation}
    x=\tau^4u\,.
    \label{eq:energydensity}
\end{equation}
Recall that the parameter~$\tau$ may be considered as an estimate of the sampling duration, so we now reintroduce~$\tau$ to account for variable measurement times.  In Sec.~\ref{sec:fit}, we found numerical estimates for the exponents in Eq.~(\ref{eq:theory}) from the slopes of the data in Fig.~\ref{fig:fit}, assuming~$\tau=1$.  The proportionality constant, which is not well predicted theoretically, may also be numerically estimated from the vertical intercepts of the data in those figures.  Accounting for the scaling factor~$R/\pi$, we find
\begin{equation}
\begin{aligned}
\omega_d \tau\approx
  \begin{cases}
    5.0 x_d^{0.34} &\text{worldline limit}~(x\lesssim x_*)\,,\\
    0.00051 x_d^{1.1} & \text{spacetime-averaged limit} (x\gtrsim x_*)\,.
  \end{cases}
  \end{aligned}
  \label{eq:numresult}
\end{equation}
Here we recall that~$\omega_d$ predicts the upper bound on the set of characteristic frequencies~$\omega_c$ that contribute to a fluctuation of magnitude~$x_d\approx x$.  We may estimate the transition between the two limits,~$x_*$, by the intersection of the worldline and spacetime-averaged limits,
\begin{equation}
5.0 x_*^{0.34} = 0.00051 x_*^{1.1}\,,
\end{equation}
from which we deduce
\begin{equation}
x_* \approx 1.8\times10^{5}\,.
\label{eq:xstarnumerical}
\end{equation}
This rough estimate of the transition between the two limits gives~$\ln(x_*)\approx12$, consistent with the data for~$p=3$ in Fig.~\ref{fig:data}.  We may rewrite Eq.~(\ref{eq:numresult}) in terms of the ratio~$u_d/u_*$, finding
\begin{equation}
\begin{aligned}
\omega_d\tau\approx
  \begin{cases}
    300 (u_d/u_*)^{0.34} &u_d\lesssim u_*\,,\\
    300 (u_d/u_*)^{1.1} & u_d\gtrsim u_*\,.
  \end{cases}
  \end{aligned}
  \label{eq:numerical}
\end{equation}
Here we follow Eq.~(\ref{eq:energydensity}) and define  $u_*=x_*\tau^{-4}$, and we have used the numerical estimate of~$x_*$ given by Eq.~(\ref{eq:xstarnumerical}).  Here we wish to point out a subtlety regarding our interpretation of~$\tau$.  In general, our numerical procedure for construction of the temporal sampling function does not require the switch-on time $\tau$ and the sampling duration, say~$t_0$, to be approximately equal. However, the specific functions used in this paper have $\tau \approx t_0$. A set of two timescale functions where $t_0 \gg \tau$ is possible was discussed in Sec. IID in Ref.~\cite{PhysRevA.104.012208}, where the dependence of  both $\hat{f}(\omega)$ and $P(x)$ upon the ratio $t_0/\tau$ was considered. However, no numerical studies using these more general functions have yet been performed.

Note that the value of $t_0/\tau$ is independent of the choice of units for $\tau$ itself. The latter is simply a choice of scale which does not alter the physical description. The meaningful quantities are the dimensionless variables, such as $\omega \tau$ of $x = \tau^4 u$.
 We argue that our focus on a subset of temporal sampling functions should not be a source of worry and may in fact be necessary in certain cases.  The correspondence between the measurement process and the sampling functions is not always clear, though we expect some relationship between the two. 
 
We now return to the question of reflection in a metal cavity.  Such reflection requires the angular frequencies of the photons to be less than the plasma frequency of the metal,~$\omega_p$.  From Eq.~(\ref{eq:numerical}), we may then find a constraint on the sampling time given a fluctuation of magnitude~$u$.  Requiring~$\omega_c < \omega_p$, we find
\begin{equation}
\begin{aligned}
\tau\gtrsim
  \begin{cases}
    300\omega_p^{-1} (u_d/u_*)^{0.34} &u_d\lesssim u_*\,,\\
    300\omega_p^{-1} (u_d/u_*)^{1.1} & u_d\gtrsim u_*\,.
  \end{cases}
  \end{aligned}
\end{equation}
For fixed~$u/u_*$, the constraint~$\omega_c < \omega_p$ thus implies that an increased~$\tau$ results in a decreased~$\omega_c$.  That is to say, a measurement over longer timescales will observe contributions from smaller frequencies, consistent with our intuition that higher frequency modes are suppressed in these cases.  We end our discussion with a specific example.  For aluminum, the plasma frequency is~$\omega_p\approx15$ eV~\cite{G_rard_2014}.  Recalling that in units where the reduced Planck constant is set to unity, 1 eV$^{-1} = 0.66$ fs, we find 
\begin{equation}
\begin{aligned}
\tau\gtrsim
  \begin{cases}
    13 \,(15\text{ eV}/\omega_p) (u_d/u_*)^{0.34} \text{ fs}&u_d\lesssim u_*\,,\\
    13 \,(15\text{ eV}/\omega_p) (u_d/u_*)^{1.1} \text{ fs} & u_d\gtrsim u_*\,.
  \end{cases}
  \end{aligned}
  \label{eq:aluminum}
\end{equation}
This result may be used to estimate the timescales in which the most of the contributing particle frequencies are small enough to allow reflection from boundaries of an aluminum cavity.  Note that because $\omega_d$ in Eq.~(\ref{eq:numerical})is an upper bound on the characteristic frequencies, timescales shorter than the bounds in Eq.~(\ref{eq:aluminum}) may still allow reduced levels of reflection.  Practical application of these results will require an estimate of~$u/u_*$ and may depend on the particular context of the experiment.  Here we merely note that because~$u$ and~$x$ differ only by a scaling constant, Eq.~(\ref{eq:energydensity}), the ratios~$u/u_*$ and~$x/x_*$ are equivalent, so  numerical simulations of the sort explored in this paper may be used directly for these estimates. Although our numerical calculations were performed assuming quantization in a spherical cavity, the analytic arguments given in Sec.~\ref{sec:theory} suggest that $\omega_d$ is independent of the cavity geometry.

 A different physical application of the characteristic frequencies involves the effects of vacuum radiation pressure fluctuations on Rydberg atoms~\cite{PhysRevA.104.012208}. These fluctuations involve an operator with~$p=7$, as shown in Ref.~\cite{Huang:2016kmx}.  The model proposed in Ref.~\cite{PhysRevA.104.012208} involves the excitation of an atom to a highly excited Rydberg state by a laser pulse, and its subsequent deexcitation by a second pulse.  The combined effect can be viewed as a measurement of the~$p=7$ operator with a two-timescale temporal sampling function, resulting in velocity fluctuations of the atom.  Upper bounds on the characteristic frequency~$\omega_c$ associated with a fluctuation of magnitude~$x$ is predicted by Eq.~(\ref{eq:logtheory}) to scale as~$x^{1/7}$ in the worldline limit. Ideally, we would like to have~$\omega_c$ small compared to the energy level separations of the atom, for which a low upper bound is a sufficient but not necessary condition.  Whether this is the case will require a more detailed analysis of involving a two-timescale sampling function, which will be a topic for future research.

\section{Conclusion}
\label{sec:conclusion}
Phenomena associated with a full theory of quantum gravity may be explored through extensions to the semiclassical theory of gravity.  Many such phenomena have been proposed over the years, including geodesic focusing~\cite{Borgman:2003dm,PhysRevLett.107.021303,PhysRevD.102.126018}, impacts on power spectra~\cite{Ford:2010wd,PhysRevD.95.063524}, increased barrier penetration rates of charged particles~\cite{Huang:2016kmx}, alternative false vacuum decay pathways for self-coupled scalar fields~\cite{PhysRevD.105.085025}, velocity fluctuations of Rydberg atoms~\cite{PhysRevA.104.012208}, and large variations in scattered photons in low-temperature light scattering experiments~\cite{PhysRevResearch.2.032028}.  Many of these effects are not gravitational and may be interesting in their own right.

Thus far, experiments searching for these effects remain lacking.  To provide future proposals with the machinery for more accurate calculations, we investigated the particle frequencies driving large vacuum quantum fluctuations of the normal ordered square of time derivatives of the massless scalar field.  We focused on the case where these fluctuations are observed in a finite duration of time and finite region of space, which we speculate is more relevant for experimental purposes.  We discovered that these quantum fluctuations are composed of a variety of particles at different angular frequencies, and the corresponding frequency spectra display complex behavior that remains to be understood.  We extracted a characteristic frequency from each spectra to assign to each quantum state, assumed to be representative of the particle frequencies contributing to the fluctuation associated with that state.  In doing so, we found that the characteristic frequencies and fluctuation magnitudes are related by a power law behavior with two distinct limits, the worldline and spacetime-averaged limits.  Furthermore, the numerical results obtained here are in good agreement with the analytic predictions, as illustrated in Table~\ref{tab:fit}.  These results allow us to describe the particle frequencies that substantially contribute to different fluctuations.  We described a simple physical application with photons arising from energy density fluctuations reflecting off the boundary of an aluminum cavity.  We showed that the measurement duration for which a majority photons are reflected is bounded below.  Such constraints may be used to infer the viability of experiments probing large quantum fluctuations.

Further investigations in this direction may be of interest, some of which we briefly outline here.  The qualitatively different behavior of the characteristic frequencies for the~$p=3$ case remains to be addressed.  Perhaps the effects of other frequency modes, such as those where~$l\neq0$ and~$m\neq0$, are more crucial.  Missing in our analysis is also any notion of probabilities: there is no guarantee that the upper bound on the characteristic frequencies is approximately equal to the most likely characteristic frequency, and the latter may also be of theoretical interest.  Analysis of the most likely frequency can be done by revising the method in Secs.~\ref{sec:omegac} and~\ref{sec:fit} to incorporate the probabilities~$P(x)$ of measuring a fluctuation of magnitude~$x$.  The connection to the probability distribution may run deeper than we considered in this paper.  The transition from the worldline to spacetime-averaged limit, Eq.~(\ref{eq:xstar}), is expected to be identical for both the probability distribution, Eq.~(\ref{eq:prob}), and the envelope of the characteristic frequencies, Eq.~(\ref{eq:theory}).  This equivalence is not guaranteed, and any discrepancies between the probability and frequency data could provide promising avenues to further understand the physics behind large quantum fluctuations.

\section{Acknowledgments}
We would like to thank Christopher Fewster
for reading and commenting upon the manuscript.  This work was supported by the National Science Foundation under Grant No. PHY-2207903. Portions of this work were conducted in the Department of Physics at the University of Jyv$\ddot{\text{a}}$skyl$\ddot{\text{a}}$, Finland, and supported in part by the Academy of Finland grant 318319.  E.D.S. is supported in part by the National Science Foundation under Award No. PHY-2114024. The authors acknowledge the Tufts University High Performance Compute Cluster (https://it.tufts.edu/high-performance-computing) which was utilized for the research reported in this paper.\\
$^\dagger$\href{mailto:Peter.Wu610348@tufts.edu}{Peter.Wu610348@tufts.edu}\\
$^\ddagger$\href{mailto:ford@cosmos.phy.tufts.edu}{ford@cosmos.phy.tufts.edu}\\
$^*$\href{mailto:enrico.d.schiappacasse@rice.edu}{enrico.d.schiappacasse@rice.edu}\\

\bibliography{FrequencySpectra} 
\end{document}